\def\Box{\kern1pt\vbox{\hrule height 1.2pt\hbox{\vrule width 1.2pt\hskip 3pt
   \vbox{\vskip 6pt}\hskip 3pt\vrule width 0.6pt}\hrule height 0.6pt}\kern1pt}
\def\gtwid{\mathrel{\raise.3ex\hbox{$>$\kern-.75em\lower1ex\hbox{$\sim$}}}}
\def\ltwid{\mathrel{\raise.3ex\hbox{$<$\kern-.75em\lower1ex\hbox{$\sim$}}}}
\def\Box{\kern1pt\vbox{\hrule height 1.2pt\hbox{\vrule width 1.2pt\hskip 3pt
   \vbox{\vskip 6pt}\hskip 3pt\vrule width 0.6pt}\hrule height 0.6pt}\kern1pt}
\documentstyle[epsfig,12pt]{article}

\begin{document}
\begin{titlepage}
\begin{flushright}
hep-th/0006207 \\ UFIFT-HEP-00-16
\end{flushright}
\vspace{.4cm}
\begin{center}
\textbf{A Canonical Formalism For Lagrangians \\
With Nonlocality Of Finite Extent}
\end{center}
\begin{center}
R. P. Woodard$^{\dagger}$
\end{center}
\begin{center}
\textit{Department of Physics \\ University of Florida \\ 
Gainesville, FL 32611}
\end{center}
\begin{center}
ABSTRACT
\end{center}
\hspace*{.5cm} I consider Lagrangians which depend nonlocally in time 
but in such a way that there is no mixing between times differing by more than 
some finite value $\Delta t$. By considering these systems as the limits of 
ever higher derivative theories I obtain a canonical formalism in which the 
coordinates are the dynamical variable from $t$ to $t + {\Delta t}$. A simple 
formula for the conjugate momenta is derived in the same way. This formalism 
makes apparent the virulent instability of this entire class of nonlocal 
Lagrangians. As an example, the formalism is applied to a nonlocal analog of 
the harmonic oscillator.
\begin{flushleft}
\end{flushleft}
\vspace{.4cm}
\begin{flushleft}
$^{\dagger}$ e-mail: woodard@phys.ufl.edu
\end{flushleft}
\end{titlepage}

\section{Introduction}

The traditional goal of fundamental physics is to infer the rules by which the
``present state'' of a system's dynamical variables determines their future 
state. Since Newton's time, most attention has been given to models for which
the ``present state'' of a system's dynamical variables means their values at
some instant in time and possibly also the values of their first time 
derivatives. This restriction corresponds to equations of motion which are 
local in time and contain no more than second time derivatives. It has not so
far proved useful, in describing the physical universe on the most fundamental
level, to invoke equations of motion which are either nonlocal in time or which
even possess more than two time derivatives. 

The deep reason behind this surprising simplification of fundamental theory 
seems to be the result obtained by the 19th century physicist Ostrogradski 
\cite{Ost}. He showed that Lagrangians which possess a finite number of higher
time derivatives, and are not degenerate in the highest one, must give rise to 
Hamiltonians which are {\it linear} in essentially half of the canonical 
variables. This is a nonperturbative result. Further, it cannot be altered by
quantization since the instability occurs over a large volume of the canonical
phase space. I will review Ostrogradski's construction in Section 2 of this 
paper. For now it suffices to note that the instability must apply as well to 
nonlocal theories which can be represented as the limits of ever higher 
derivative ones. 

Much of the interest in nonlocal quantum field theories has been driven by the
close connection between ultraviolet divergences and local interactions 
\cite{Pais}. Of course it does no good to avoid divergences by introducing an 
infinite number of instabilities against which there is not even any barrier to
decay! It is therefore of interest to know when a nonlocal Lagrangian possesses
a higher derivative representation and, consequently, the Ostrogradskian 
instability. The higher derivative representation does seem to be valid for 
cases such as string field theory, where the nonlocality enters through entire 
functions of the derivative operator and the Lagrangian cannot be made local by
a field redefinition \cite{Eliezer}. On the other hand, the higher derivative
representation is certainly not valid for the inverse differential operators
which result from integrating out a local field variable. It also fails for
``maximal nonlocality'' in which the action is a nonlinear function of local 
actions \cite{Bennett}. The purpose of this paper is to demonstrate by 
construction that the higher derivative representation is valid for 
``nonlocality of finite extent'' in which the Lagrangian connects no times 
differing by more than some constant ${\Delta t}$.

Although the results of this paper apply as well to field theories I will work 
in the context of a one dimensional, point particle whose position as a 
function of time is $q(t)$. A nonlocal Lagrangian of finite extent ${\Delta t}$
is one which definitely depends upon (and mixes) $q(t)$ and $q(t +{\Delta t})$,
and potentially depends as well upon $q(t')$ for $t < t' < t + {\Delta t}$. An 
example would be the following nonlocal generalization of the harmonic 
oscillator:
\begin{equation}
L[q](t) = \frac12 m \dot{q}^2(t + {\Delta t}/2) - \frac12 m \omega^2 q(t) q(t +
{\Delta t}) \; . \label{eq:exampL}
\end{equation}
The deterministic way of viewing such theories is that the equations of motion
give the dynamical variable at the latest time --- $q(t+{\Delta t})$ --- as a
function of earlier times in the range $t-{\Delta t} \leq t' < t+{\Delta t}$. 
In our example the equation of motion is
\begin{equation}
\int_0^{\Delta t} dr {\delta L[q](t-r) \over \delta q(t)} = - m \left\{\ddot{q
}(t) + \frac12 \omega^2 q(t + {\Delta t}) + \frac12 \omega^2 q(t- {\Delta t})
\right\} = 0 \; ,
\end{equation}
and its deterministic interpretation is
\begin{equation}
q(t+{\Delta t}) = - q(t-{\Delta t}) - \frac{2}{\omega^2} \ddot{q}(t) \; .
\end{equation}

This paper contains six sections of which this Introduction is the first. 
Section 2 is devoted to a review of Ostrogradski's result for local Lagrangians 
depending upon $N$ time derivatives. My canonical formalism is presented in 
Section 3 and shown to correctly realize the dynamics of nonlocal Lagrangians
of finite extent. This formalism is applied in Section 4 to the Lagrangian 
(\ref{eq:exampL}) discussed above. The connection with Ostrogradski's formalism
is demonstrated in Section 5. My conclusions comprise Section 6.

\section{Ostrogradski's construction}

Consider a Lagrangian $L\left(q,\dot{q},\dots,q^{(N)}\right)$ which depends 
upon the first $N$ derivatives of the dynamical variable $q(t)$. I shall assume
only that the Lagrangian is {\it nondegenerate}, i.e., that the equation
\begin{equation}
P_N = {\partial L \over \partial q^{(N)}} \; ,
\end{equation}
can be inverted to solve for $q^{(N)}$ as a function of $P_N$, $q$ and the 
first $N-1$ derivatives of $q$. This just means that the action's dependence 
upon $q^{(N)}$ cannot be eliminated by partial integration, so the equation of
motion,
\begin{equation}
\sum_{I=0}^N \left(-{d \over dt}\right)^I {\partial L \over \partial q^{(I)}}
= 0 \; , \label{eq:EOM}
\end{equation}
contains $q^{(2N)}$. 

Since the equation of motion determines $q^{(2N)}$ as a function of $q$ and its
first $2N-1$ derivatives, one can obviously specify the initial values of these
$2N$ variables. The canonical phase space must accordingly contain $N$ 
coordinates and $N$ conjugate momenta. In Ostrogradski's construction 
\cite{Ost} the $I$-th coordinate is just the $(I-1)$-th derivative of $q$,
\begin{equation}
Q_I \equiv q^{(I-1)} \; . \label{eq:QI}
\end{equation}
The momentum canonically conjugate to $Q_I$ is,
\begin{equation}
P_I = \sum_{J=I}^N \left(-{d \over dt}\right)^{J-I} {\partial L \over \partial
q^{(J)}} \; . \label{eq:PI}
\end{equation}
A consequence of nondegeneracy is that the derivatives $q^{(N+I)}$ can be 
determined from $P_{N-I},P_{N-I+1},\dots,P_N$ and the $Q_J$'s. In particular 
$q^{(N)}$ involves only $P_N$ and the $Q_J$'s,
\begin{equation}
q^{(N)} = {\cal Q}\left(\vec{Q},P_N\right) \; . \label{eq:QN}
\end{equation}

Ostrogradski's Hamiltonian is,
\begin{eqnarray}
H & = & \sum_{I=1}^N P_I \dot{Q}_I - L \; , \\
& = & \sum_{I=1}^{N-1} P_I Q_{I+1} + P_N {\cal Q}\left(\vec{Q},P_N\right)
- L\left(\vec{Q},{\cal Q}\left(\vec{Q},P_N\right)\right) \; ,
\end{eqnarray}
and his canonical equations are the ones suggested by the notation,
\begin{equation}
\dot{Q}_I = {\partial H \over \partial P_I} \qquad , \qquad \dot{P}_I = - 
{\partial H \over \partial Q_I} \; .
\end{equation}
It is straightforward to check that the various canonical evolution equations 
reproduce the equation of motion and the structure of the canonical formalism:
$\dot{Q}_I$ gives the canonical definition (\ref{eq:QI}) for $Q_{I+1}$; 
$\dot{Q}_N$ gives the canonical definition for $P_N$ in its inverse form 
(\ref{eq:QN}); $\dot{P}_{I+1}$ gives the canonical definition (\ref{eq:PI}) for
$P_I$; and $\dot{P}_1$ gives the equation of motion (\ref{eq:EOM}). So there is
no doubt that Ostrogradski's Hamiltonian generates time evolution. When the
Lagrangian is free of explicit time dependence $H$ is also the conserved 
current associated with time translation invariance.

The instability consequent upon $H$'s linearity in $P_1,P_2,\dots,P_{N-1}$ 
explains why higher derivative theories have not been of use in describing 
physics on the fundamental level. Note the generality of the problem. It does
not depend upon any approximation scheme, nor upon any feature of the 
Lagrangian except nondegeneracy. Further, it must continue to afflict the
theory after quantization because the instability is not confined to a small
region of the classical phase space. If a fully nonlocal Lagrangian can be
represented as the limit of such higher derivative Lagrangians it must inherit
their instability.

The limit of infinite $N$ is facilitated by regarding Ostrogradski's formalism
as the result of constraining a larger system with an extra pair of canonical
variables,
\begin{equation}
Q_{N+1} \equiv q^{(N)} \qquad , \qquad P_{N+1} \approx 0 \; .
\end{equation}
The Hamiltonian is,
\begin{equation}
H = \sum_{I=1}^N P_I Q_{I+1} - L\left(\vec{Q},Q_{N+1}\right) \; ,
\end{equation}
and requiring that $P_{N+1}$ remains zero imposes the canonical definition of 
$P_N$ as another constraint,
\begin{equation}
\dot{P}_{N+1} = - {\partial H \over \partial Q_{N+1}} = - P_N + {\partial L
\over Q_{N+1}} \approx 0 \; .
\end{equation}
Since the Poisson bracket with $P_{N+1}$ gives the second derivative of the 
Lagrangian with respect to $Q_{N+1}$, nondegeneracy implies that the two
constraints are second class. The resulting Dirac brackets are,
\begin{eqnarray}
\left\{Q_I,Q_J\right\}_D & = & \left(-\delta_{I \; N+1} \delta_{J \; N} +
\delta_{I \; N} \delta_{J \; N+1}\right) \left[{\partial^2 L \over \partial 
Q^2_{N+1}}\right]^{-1} \; , \\
\left\{Q_I,P_J\right\}_D & = & \delta_{I \; J} - \delta_{I \; N+1} \left[{
\partial^2 L \over \partial Q^2_{ N+1}}\right]^{-1} {\partial^2 L \over 
\partial Q_J \partial Q_{N+1}} \; , \\
\left\{P_I,P_J\right\}_D & = & 0 \; .
\end{eqnarray}
Note that there is not even any difference between Dirac brackets and Poisson 
brackets provided one avoids the highest $Q$ --- that is, $Q_{N+1}$.

\section{My construction for finite nonlocality}

I define a nonlocal Lagrangian $L[q](t)$ of finite extent ${\Delta t}$ as one
which potentially depends upon the dynamical variable from time $t$ to time
$t+ {\Delta t}$, with guaranteed mixing between $q(t)$ and $q(t+{\Delta t})$. 
The requirement of mixing is the generalization of nondegeneracy and it 
implies,
\begin{equation}
{\delta^2 L[q](t) \over \delta q(t) \delta q(t+{\Delta t})} \neq 0 \; .
\end{equation}
I shall also require that the Lagrangian contain no derivatives of either
$q(t)$ or $q(t+{\Delta t})$.

I label the canonical variables by a continuum parameter $0 \leq s \leq {\Delta
t}$. They are defined as follows:
\begin{eqnarray}
Q(s,t) & \equiv & q(s+t) \; , \label{eq:myQ} \\ 
P(s,t) & \equiv & \int_s^{\Delta t} dr {\delta L[q](s+t-r) \over \delta q(s+t)
\quad} \; . \label{eq:myP}
\end{eqnarray}
Note that (\ref{eq:myP}) implies the constraint $P({\Delta t},t) \approx 0$.
Note also that whereas the $s$ and $t$ derivatives of $Q(s,t)$ are identical 
those of $P(s,t)$ are not,
\begin{equation}
{d \over ds} P(s,t) = {d \over dt} P(s,t) - {\delta L[q](t) \over \delta 
q(s+t)} \; . \label{eq:Pdifs}
\end{equation}
Since $L[q](t)$ involves the dynamical variable from $q(t)$ up to $q(t+{\Delta 
t})$ we see that $P(s,t)$ involves $q(s+t-{\Delta t})$ up to $q(t+{\Delta t})$.
So decreasing $s$ allows one to reach back further before time $t$, all the way
to time $t-{\Delta t}$ at $s=0$.

Note that the equation of motion is $P(0,t) = 0$. This emerges as an additional
constraint from surface variations of the canonical Hamiltonian,
\begin{equation}
H(t) \equiv \int_0^{\Delta t} dr P(r,t) {d \over dr} Q(r,t) - L[Q](t) \; .
\end{equation}
We can find the canonical equations of time evolution from the fact that the
only nonzero Poisson bracket is,
\begin{equation}
\left\{Q(r,t),P(s,t)\right\} = \delta(r-s) \; . \label{eq:com}
\end{equation}
The result for $Q(s,t)$ is straightforward,
\begin{eqnarray}
{d \over dt} Q(s,t) & = &\left\{Q(s,t),H(t)\right\} \; , \\
& = & \int_0^{\Delta t} dr \delta(r-s) {d \over dr} Q(r,t) \; , \\
& = & {d \over ds} Q(s,t) \; .
\end{eqnarray}
A partial integration is necessary for $P(s,t)$ and one must be careful about
the resulting surface terms,
\begin{eqnarray}
{d \over dt} P(s,t) & = & \left\{P(s,t),H(t)\right\} \; , \\
& = & - \int_0^{\Delta t} dr P(r,t) {d \over dr} \delta(r - s) +{\delta L[Q](t)
\over \delta Q(s,t)} \; , \\
& = & {d \over ds} P(s,t) + {\delta L[Q](t) \over \delta Q(s,t)} - \delta(r-s) 
P(r,t) {\left\vert {\mbox{} \over \mbox{}} \right.}_0^{\Delta t} \; . 
\label{eq:dP/dt}
\end{eqnarray}
For $0 < s < {\Delta t}$ this simply reproduces (\ref{eq:Pdifs}), and hence the
canonical definition of $P(s,t)$. 

Since the two surface terms cannot be canceled by anything else, they must 
be imposed as constraints,
\begin{equation}
P(s,t) \approx 0 \qquad s = 0, {\Delta t} \; .
\end{equation}
Requiring that they be preserved under time evolution implies two additional
constraints,
\begin{equation}
{d \over ds} P(s,t) + {\delta L[Q](t) \over \delta Q(s,t)} \approx 0 \qquad
s = 0, {\Delta t} \, .
\end{equation}
Nondegeneracy --- and the absence in $L[q](t)$ of derivatives of $q(t)$ and/or 
$q(t+{\Delta t})$ --- guarantees that the four constraints are second class.

Note that the $H(t)$ is conserved when $L[q](t)$ is free of explicit time
dependence,
\begin{eqnarray}
{d H\over dt}(t) & = & \int_0^{\Delta t} ds \left\{ {dP \over dt}(s,t)
{dQ \over ds}(s,t) + P(s,t) {d^2 Q \over ds^2}(s,t)\right\} -
{d \over dt} L[Q](t) \; , \\
& = & \int_0^{\Delta t} ds {d \over ds} \left[P(s,t) {d \over ds} Q(s,t)
\right] \nonumber \\
& & \qquad \qquad + \int_0^{\Delta t} ds {\delta L[Q](t) \over \delta Q(s,t)} 
{d \over ds} Q(s,t) - {d \over dt} L[Q](t) \; , \\
& = & \left. P(s,t) {d \over ds} Q(s,t) \right\vert_0^{\Delta t} \; , \\
& \approx & 0 \; .
\end{eqnarray}
Note also that the Hamiltonian has inherited the Ostrogradskian instability.
After eliminating the constraints it must be linear in all the $P(s,t)$
except possibly ${d \over ds} P(s,t)$ at $s=0$ and at $s={\Delta t}$.

\section{A simple example}

It is useful to see how the general construction given in the previous section
applies to the Lagrangian (\ref{eq:exampL}) presented in Section 1. Of course
the canonical coordinates are always $Q(s,t) = q(s+t)$ for $0 \leq s \leq 
{\Delta t}$. To find the canonical momenta note that the functional derivative 
of the Lagrangian is,
\begin{eqnarray}
{\delta L[q](t) \over \delta q(s+t)} & = & -m \dot{q}\left(t + \frac{\Delta t}2
\right) \delta'\left(s - \frac{\Delta t}2\right) \nonumber \\
& & \qquad - \frac12 m \omega^2 \left[{\mbox{} \over \mbox{}} q(t + {\Delta t})
\delta(s) + q(t) \delta(s - {\Delta t})\right] \; .
\end{eqnarray}
Substituting in (\ref{eq:myP}) gives,
\begin{eqnarray}
P(s,t) & = & m \dot{q}\left(t + \frac{\Delta t}2\right) \delta\left(s - \frac{
\Delta t}2\right) - m \ddot{q}(s+t) \theta\left(\frac{\Delta t}2 - s\right) 
\nonumber \\
& & \; - \frac12 m \omega^2 \left[{\mbox{} \over \mbox{}} q(s + t + {\Delta 
t}) \theta(-s) + q(s + t -{\Delta t}) \theta({\Delta t} - s)\right] \; . \qquad
\label{eq:Pex}
\end{eqnarray}
Note that $P({\Delta t},t) = 0$ and that
\begin{equation}
P(0,t) = -m \left[\ddot{q}(t) + \frac12 \omega^2 q(t + {\Delta t}) + \frac12
\omega^2 q(t - {\Delta t})\right] \; , \label{eq:exEOM}
\end{equation}
indeed vanishes with the equation of motion.

The canonical Hamiltonian is
\begin{equation}
H(t) = \int_0^{\Delta t} ds P(s,t) {d Q\over ds}(s,t) - \frac12 m \left[{d Q
\over ds}\left(\frac{\Delta t}2,t\right)\right]^2 + \frac12 m \omega^2 Q(0,t) Q({\Delta 
t},t) \; .
\end{equation}
The canonical evolution equations are,
\begin{eqnarray}
{d Q \over dt}(s,t) & = & {d Q \over ds}(s,t) \; ,\\
{d P \over dt}(s,t) & = & {d P \over ds}\left(s,t\right) - m {d Q \over 
ds}\left(\frac{\Delta t}2,t\right) \delta'\left(s - \frac{\Delta t}2\right) 
\nonumber \\
& & \qquad - \frac12 m \omega^2 \left[Q({\Delta t},t) \delta(s) + Q(0,t)
\delta(s - {\Delta t})\right] \; .
\end{eqnarray}
It is simple to check that substituting $Q(s,t) = q(s+t)$ and relation 
(\ref{eq:Pex}) for $P(s,t)$ indeed verifies these equations.

The constraints are $P(0,t) \approx 0$, $P({\Delta t},t) \approx 0$ and the
apparently singular pair,
\begin{eqnarray}
{d P \over ds}(0,t) - \frac12 m \omega^2 Q({\Delta t},t) \delta(0) & \approx &
0 \; , \\
{d P \over ds}({\Delta t},t) - \frac12 m \omega^2 Q(0,t) \delta(0) & \approx &
0 \; .
\end{eqnarray}
However, the vanishing of $P(s,t)$ at the endpoints means that the endpoint 
derivatives contain delta functions, so the actual constraints are the 
perfectly regular coefficients of $\delta(0)$,
\begin{eqnarray}
P(0^+,t) - \frac12 m \omega^2 Q({\Delta t},t) & \approx & 0 \; , \\
- P({\Delta t}^-,t) - \frac12 m \omega^2 Q(0,t) & \approx & 0 \; .
\end{eqnarray}
Note that these constraints are implied by (\ref{eq:Pex}) and, where necessary,
the vanishing of (\ref{eq:exEOM}). Note also that the contstraints determine
both the actual endpoint values of $P(s,t)$ and its limit as the endpoints are
approached.

Since the Lagrangian (\ref{eq:exampL}) has no explicit dependence upon time the
Hamiltonian should be conserved. To see that it is, first substitute $Q(s,t) =
q(s,t)$ and relation (\ref{eq:Pex}) for $P(s,t)$ to obtain,
\begin{eqnarray}
\lefteqn{\int_0^{\Delta t} ds P(s,t) {d Q \over dq}(s,t) = m \dot{q}^2\left(t + 
\frac{\Delta t}2 \right) - m \int_0^{{\Delta t}/2} ds \dot{q}(s+t) \ddot{q}(s
+ t)} \nonumber \\
& & \qquad \qquad - \frac12 m \omega^2 \int_0^{\Delta t} ds q(s+t-{\Delta t}) 
\dot{q}(s+t) \; , \\
& & = \frac12 m \left[\dot{q}^2\left(t + \frac{\Delta t}2\right) + \dot{q}^2(t)
\right] - \frac12 m \omega^2 \int_0^{\Delta t} ds q(s + t - {\Delta t}) 
\dot{q}(s + t) \; . \qquad
\end{eqnarray}
Then subtract expression (\ref{eq:exampL}) to determine the configuration space
Hamiltonian,
\begin{equation}
H(t) = \frac12 m \dot{q}^2(t) + \frac12 m \omega^2 q(t) q(t + {\Delta t}) -
\frac12 m \omega^2 \int_0^{\Delta t} ds q(s + t - {\Delta t}) \dot{q}(s + t)
\; .
\end{equation}
Now use the fact that the integrand depends upon $t$ only through the sum $s+t$
to express the derivative of the integral as a surface term,
\begin{eqnarray}
{d H \over dt}(t) & = & m \dot{q}(t) \ddot{q}(t) + \frac12 m \omega^2 \left[
\dot{q}(t) q(t + {\Delta t}) + q(t) \dot{q}(t + {\Delta t})\right] \nonumber \\
& & \qquad - \left.  \frac12 m \omega^2 q(s + t - {\Delta t}) \dot{q}(s + t) 
\right\vert_{s = 0}^{s = {\Delta t}} \; , \\
& = & m \dot{q}(t) \left[\ddot{q}(t) + \frac12 \omega^2 q(t - {\Delta t}) +
\frac12 \omega^2 q(t + {\Delta t})\right] \; .
\end{eqnarray}

The most straightforward way of demonstrating that the transformation to the 
constrained phase space is invertible is by exhibiting the inverse. Of course 
we always have,
\begin{equation}
q(s+t) = Q(s,t) \qquad , \qquad \forall \; 0 \leq s \leq {\Delta t} \; .
\end{equation}
For $-{\Delta t} < s < 0$ one recovers $q(s+t)$ from relation (\ref{eq:Pex}),
\begin{equation}
q(s+t) = -{2 \over m \omega^2} P(s+{\Delta t},t) - {2 \over \omega^2} {d \over
ds} \left[{dQ \over ds}(s+{\Delta t},t) \theta\left(-\frac{\Delta t}2 - s
\right) \right] \; .
\end{equation}
The endpoint case of $s = - {\Delta t}$ is given by the constraint $P(0,t) 
\approx 0$,
\begin{equation}
q(t - {\Delta t}) = - Q({\Delta t},t) - {2 \over \omega^2} {d^2 Q \over 
ds^2}(0,t) \; ,
\end{equation}
where I am of course defining differentiation in the right-handed sense,
\begin{equation}
{d f \over dx}(x) \equiv \lim_{\epsilon \rightarrow 0^+} {f(x+\epsilon) - f(x)
\over \epsilon} \; .
\end{equation}

It is amusing to close the section by exhibiting the run-away solutions which 
are one possible consequence of the Ostrogradskian instability. Since the 
configuration space equation of motion,
\begin{equation}
\ddot{q}(t) + \frac12 \omega^2 [q(t + {\Delta t}) + q(t - {\Delta t})] = 0 \; ,
\end{equation}
is linear and invariant under time translation, the general solution must be a
superposition of terms having the form $e^{ikt}$. The allowed frequencies are 
complex numbers $k$ which obey,
\begin{equation}
k^2 = \omega^2 \cos(k {\Delta t}) \; .
\end{equation}
The equation is transcendental but graphing both sides shows a single pair of
$\pm$ real solutions. To find the remaining solutions make the substitution,
\begin{equation}
k = \alpha + i \beta \; ,
\end{equation}
and take the real and imaginary parts of the equation,
\begin{eqnarray}
\alpha^2 - \beta^2 & = & \omega^2 \cos(\alpha {\Delta t}) \cosh(\beta {\Delta 
t}) \; ,\\
2 \alpha \beta & = & - \omega^2 \sin(\alpha {\Delta t}) \sinh(\beta {\Delta t})
\; .
\end{eqnarray}
Graphical analysis indicates a conjugate pair of solutions for $\alpha {\Delta
t}$ in each $2\pi$ interval of the real line. For large integer $N$ these 
solutions have the form,
\begin{eqnarray}
\alpha {\Delta t} & \approx & 2 \pi N - {2 \ln(N) \over \pi N} \; , \\
\pm \beta {\Delta t} & \approx & \ln\left({8 \pi^2 N^2 \over \omega^2 {\Delta 
t}^2}\right) + \left({\ln(N) \over \pi N}\right)^2 \; .
\end{eqnarray}
So this system has the infinite number of solutions predicted by the 
Ostrogradskian analysis, and all but two them grow or fall exponentially.

\section{Ostrogradskian derivation}

My representation is related to the infinite $N$ limit of Ostrogradski's 
through the Maclaurin series,
\begin{equation}
Q(s,t) = \sum_{I=0}^{\infty} {s^I \over I!} Q_{I+1}(t) \; .
\end{equation}
Note that differentiation with respect to the Ostrogradskian coordinates is 
realized by the functional chain rule,
\begin{equation}
{\partial \qquad \over \partial Q_I(t)} = \int_0^{\Delta t} ds \left[{\partial 
Q(s,t) \over \partial Q_I(t)}\right] {\delta \qquad \over \delta Q(s,t)} = 
\int_0^{\Delta t} ds {s^{I-1} \over (I-1)!} {\delta \qquad \over \delta Q(s,t)} 
\; ,
\end{equation}
where the functional derivative is defined by
\begin{equation}
{\delta Q(r,t) \over \delta Q(s,t)} = \delta(r-s) \; ,
\end{equation}
and the ordinary rules of calculus. From (\ref{eq:myQ}) one obtains a useful
formula for the higher derivative representation,
\begin{equation}
{\partial L[q](t) \over \partial q^{(I)}(t)} = \int_0^{\Delta t} ds {s^I \over 
I!} {\delta L[q](t) \over \delta q(s+t)} \; .
\end{equation}

The conjugate momentum $P(s,t)$ should depend linearly on the Ostrogradskian 
momenta,
\begin{equation}
P(s,t) = \sum_{I=0}^{\infty} p_I(s) P_{I+1}(t) \; . \label{eq:linrel}
\end{equation}
The combination coefficients $p_I(s)$ can be determined by enforcing the 
canonical Poisson bracket (\ref{eq:com}),
\begin{eqnarray}
\delta(r-s) & = & \sum_{I=0}^{\infty} {r^I \over I!} \sum_{J=0}^{\infty} p_J(s)
\left\{ Q_{I+1}(t),P_{J+1}(t)\right\} \; , \\
& = & \sum_{I=0}^{\infty} {r^I \over I!} p_I(s) \; .
\end{eqnarray}
By acting $(\partial/{\partial r})^J$ and then taking $r \rightarrow 0$ one
finds,
\begin{equation}
p_J(s) = \left(-{d \over ds}\right)^J \delta(s) \; . \label{eq:coefs}
\end{equation}

To obtain my formula (\ref{eq:myP}) for the conjugate momenta note first that,
for infinite $N$, the Ostrogradskian momenta are,
\begin{eqnarray}
P_I(t) & = & \sum_{J=I}^{\infty} \left(-{d \over dt}\right)^{J-I} {\partial 
L[q](t) \over \partial q^{(J)}(t)} \; , \\
& = & \sum_{J=I}^{\infty} \left({d \over dt}\right)^{J-I} \int_0^{\Delta t} dr
{r^J \over J!} {\delta L[q](t) \over \delta q(r+t)} \; .
\end{eqnarray}
Now substitute this and (\ref{eq:coefs}) into (\ref{eq:linrel}),
\begin{equation}
P(s,t) = \sum_{I=0}^{\infty} \left[\left(-{d \over ds}\right)^I \delta(s)
\right] \sum_{J = I+1}^{\infty} \left(-{d \over dt}\right)^{J-I-1} \int_0^{
\Delta t} dr {r^J \over J!} {\delta L[q](t) \over \delta q(r+t)} \; .
\end{equation}
Simplification is achieved by exploiting the identity,
\begin{equation}
{r^J \over J!} = \int_0^r dr' {(r-r')^I (r')^{J-I-1} \over I! (J-I-1)!} \; ,
\end{equation}
to recognize the two sums as Taylor expansions of the shift operator,
\begin{eqnarray}
P(s,t) & = & \int_0^{\Delta t} dr \int_0^r dr' \sum_{I=0}^{\infty} {(r-r')^I 
\over I!} \left(-{d \over ds}\right)^I \delta(s) \nonumber \\
& & \qquad \qquad \times \sum_{J=I+1}^{\infty} {(r')^{J-I-1} \over (J-I-1)!}
\left(-{d \over dt}\right)^{J-I-1} {\delta L[q](t) \over \delta q(r+t)} \; ,\\
& = & \int_0^{\Delta t} dr \int_0^r dr' \delta(s - r + r') {\delta L[q](t-r')
\over \delta q(r+t-r')} \; , \\
& = & \int_s^{\Delta t} dr {\delta L[q](s+t-r) \over \delta q(s+t)} \; .
\end{eqnarray}

The Hamiltonian follows similarly,
\begin{eqnarray}
H(t) & = & \sum_{I=1}^{\infty} P_I(t) Q_{I+1}(t) - L[Q](t) \; , \\
& = & \sum_{I=1}^{\infty} \int_0^{\Delta t} ds {s^{I-1} \over (I-1)!} P(s,t) 
\left({d \over dr}\right)^I Q(r,t) {\left\vert {\mbox{} \over \mbox{}} 
\right.}_{r=0} - L[Q](t) \; , \\
& = & \int_0^{\Delta t} ds P(s,t) {d \over ds} Q(s,t) - L[Q](t) \; .
\end{eqnarray}
Its instability is manifest from the fact that it has been derived from 
Ostrogradski's result in the limit that the number of derivatives becomes 
infinite.

\section{Discussion}

I have shown that Lagrangians with nonlocality of finite extent ${\Delta t}$ 
can be treated as the limits of higher derivative Lagrangians. I have also 
given a canonical formalism that is somewhat more natural in which the 
canonical variables are labelled by a continuum parameter $s$, for $0 \leq s 
\leq {\Delta t}$. The canonical coordinates are just the dynamical variables at
times $t+s$. A quantum mechanical state in such a system would be a functional 
of these coordinates. The conjugate momenta (\ref{eq:myP}) are given by a 
simple integral of a functional derivative of the Lagrangian. With the 
canonical coordinates, the momenta allow one to reconstruct the dynamical 
variables at times $t-s$.

There is no physical motivation for this exercise because all such models are 
virulently unstable. Indeed, the only point of the formalism is to remove any 
doubt about a possible phenomenological role for these Lagrangians. They have
inherited the full Ostrogradskian instability: essentially half of the 
directions in the classical phase space access arbitarily negative energies.
There is not even any barrier to decay. This is a nonperturbative result and,
because it arises from a large region of phase space, it must survive 
quantization.

Negative results of such power and generality are to mathematically inclined
physicists like a red flag is supposed to be to a bull. Nothing I can honestly 
add is likely to much discourage further attempts to carve out a physical niche
for nonlocal Lagrangians but I do recommend that these efforts be preceded by 
sober reflection upon the following fact: {\it in the long struggle of our
species to understand the universe it has never once proven useful to invoke a 
theory which is nonlocal on the most fundamental level.} Yet the subset of 
local Lagrangians containing no more than first derivatives is a minuscule 
fragment of the set of all functionals of the dynamical variable. The 
Ostrogradskian instability offers a simple and compelling explanation for the 
complete dominance of this tiny subset over its much larger whole. The only 
alternative would seem to be coincidence on a scale that makes even the worst 
fine tuning problem seem inconsequential.

\vskip 1cm

{\it Note Added:} Shortly after posting this paper I learned of important work
by Llosa and Vives \cite{LV} on the problem of canonically formulating a 
general nonlocal Lagrangian. My work can be viewed as a specialization of their
technique to the case of nonlocality of finite extent where the Euler-Lagrange
equations are deterministic, where an explicit Poisson bracket structure can be
determined and where the formalism can be derived from the infinite $N$ limit 
of Ostrogradski's construction. (None of these features can be present in the 
general case.) Note should also be taken of the recent work of Gomis, Kamimura 
and Llosa on canonically formulating spacetime noncommutative theories
\cite{GKL}.

\vskip 1cm
\centerline{\bf Acknowledgments}

I thank T. Jacobson for asking the question which stimulated me to carry out
this exercise. This work was partially supported by DOE contract 
DE-FG02-97ER\-41029 and by the Institute for Fundamental Theory.

\end{document}